\documentclass[11pt]{article}
\usepackage{amssymb}
\usepackage{graphics}
\usepackage{epsfig}
\usepackage{a4wide}

\begin{document}
\newcommand{\gbWWb}{$gb \to W^+W^-b$ }
\newcommand{\gbWWbj}{$gb \to W^+W^-b+j$ }
\newcommand{\gbWt}{$gb \to W^-t$ }
\newcommand{\ppgbWt}{$pp/p\bar p \to gb \to W^-t+X$ }
\newcommand{\ppWWjX}{$pp \to W^+W^-j+X$ }
\newcommand{\ppWWb}{$pp \to W^+W^-b+X$ }
\newcommand{\ppWWbb}{$pp \to W^+W^-b(\bar b)+X$ }
\newcommand{\ppWWbj}{$pp \to W^+W^-b+j+X$ }
\newcommand{\ppWWbbar}{$pp \to W^+W^-\bar{b}+X$ }
\newcommand{\gbWWbbbar}{$g\bar{b} \to W^+W^-\bar{b}$ }

\title{ QCD NLO predictions to $W$-pair production
in association with a massive (anti)bottom-jet at the LHC  }
\author{ Yan Han, Wang Shao-Ming, Ma Wen-Gan, Zhang Ren-You, and Guo Lei \\
{\small Department of Modern Physics, University of Science and Technology}  \\
{\small of China (USTC), Hefei, Anhui 230026, P.R.China}    }

\date{}
\maketitle \vskip 15mm
\begin{abstract}
The $W$-pair production in association with a massive
(anti)bottom jet is not only an important background to a number of
interesting processes, such as the single top production associated
with a W boson, but also a potential background to new physics
searches. We present the calculations of the total and differential
cross sections for the $W^+W^-+ b(\bar{b})$ jet productions at the
LHC up to the QCD next-to-leading order (NLO). Our results by
adopting the QCD NLO contribution collection scheme-I show that the
K factors can be $1.66$ and $1.21$ with the inclusive and
exclusive two-jet event selection schemes respectively, when we
set $m_H=120~GeV$, $\mu=m_W+m_b/2$ and take the constraints of
$p_{T,b(\bar{b})}>25~GeV$, $|y_{b(\bar{b})}|<2.5$ for
$b(\bar{b})$ jet. We find that the stabilization of the theoretical
prediction for the integrated cross section for the $pp \to
W^+W^-b(\bar b)+X$ up to the QCD NLO requires a veto on a second
isolated hard jet and the inclusion of the QCD NLO contribution from
the $W^+W^-b\bar{b}(bb, \bar{b}\bar{b})$ production with the final
two $b(\bar{b})$ quarks being merged as one jet.
\end{abstract}


\vfill
\eject
\baselineskip=0.32in

\renewcommand{\theequation}{\arabic{section}.\arabic{equation}}
\renewcommand{\thesection}{\Roman{section}.}
\newcommand{\nb}{\nonumber}

\newcommand{\Dir}{\kern -6.4pt\Big{/}}
\newcommand{\Dirin}{\kern -10.4pt\Big{/}\kern 4.4pt}
\newcommand{\DDir}{\kern -7.6pt\Big{/}}
\newcommand{\DGir}{\kern -6.0pt\Big{/}}

\makeatletter      
\@addtoreset{equation}{section}
\makeatother       

\section{Introduction}
\par
At the LHC, the single top associated production, $pp \to bg \to W
t+X$, offers a unique possibility of the direct measurement of the
entry $V_{tb}$ of the Cabibbo-Kobayashi-Maskawa quark-mixing matrix (CKM),
allowing nontrivial tests of the
properties of this matrix in the standard model (SM)\cite{Tait,3,4}.
By this channel we can also study the $W-t-b$ vertex, and test
precisely the V-A structure of the charged current weak interaction
of the top quark by looking at the polarization of this
quark\cite{4,Carlson}. Furthermore, such a production channel could be
interesting in hunting for new physics beyond the SM. The new
physics may manifest itself via either loop effects, or inducing
non-SM weak interactions to introduce new single top production
channels\cite{Gerber}. The produced top quark of the $p p \to b g
\to W t + X$ process subsequently decays mainly into a $W$ boson and
a bottom quark, therefore, the observed final states of the single
top-quark production in association with a $W$ boson are mostly
$W^+W^-b$ where the two $W$ bosons can be reconstructed from their
leptonic decay products. The similar analysis also applies for the
$pp \to W^+W^- \bar{b}+X$ process.

\par
Besides, the Higgs boson can be produced in association with a high
$p_T$ jet. If the SM Higgs boson is in the intermediate mass range,
the $pp \to H^0$+jet $\to W^+W^-$+jet process is expected to be a
discovery channel for the Higgs boson, particularly if the Higgs-boson
mass is very close to the threshold for the $W$-pair
production. B. Mellado, {\it et al.,} conclude in their
paper\cite{Mellado} that the Higgs signal significance can be
improved by considering also the final states which consist of the
leptonic decays of the W pair, and at least one additional jet. They
suggest that the Higgs boson associated with only a single jet at large
rapidity is requested as a signal. They find that the backgrounds
can be reduced substantially and the sensitivity of experiments can
be enhanced in the search for the Higgs boson via the $W^+ W^- +{\rm
jet}$ production channel. The QCD next-to-leading order (NLO) corrections to the process
\ppWWjX were studied in Refs.\cite{wwjet_0908.4124, wwjet
0710.1832}. In these papers the initial quark $q \neq b, \bar{b}$
was assumed, and the subprocesses with $j = b, \bar{b}$ in final
states are excluded for the reason that such processes can be excluded
by anti-$b$ tagging or should be seen as continuations of $W^-t$ or
$W^+\bar{t}$ production. Since the process \ppWWbb at the LHC is an
essential background of both the single top production and the
single Higgs-boson production channels, the investigation of it can
serve as a complementary work for those presented in
Refs.\cite{wwjet_0908.4124, wwjet 0710.1832}. Furthermore, the
process \ppWWbb is interesting in its own right, since $W$-pair
production processes enable a direct precise probing of the
nonabelian gauge boson couplings. To sum up, improving the
precision of the predictions for the $W^+W^-b(\bar b)$ production is
necessary, since $W^+W^-+b(\bar{b})$-jet production delivers not
only the background to a number of interesting processes, but also a
potential background to new physics searches.

\par
In this work, we perform the calculations for the QCD NLO
corrections to the \ppWWbb process. Within this work, the mass of
bottom quark is retained in all the partonic processes. The paper is
organized as follows. We present the details of the calculation
strategies in Sec. II. The numerical results and discussions are
given in Sec. III, and finally a short summary is given.

\vskip 5mm
\section{Strategies in calculation}\label{calc}
\par
We neglect the quark mixing between the two light and the third
massive generations (i.e., $V_{ub} = V_{cb} = V_{td} = V_{ts} \sim
0$), and take the u, d, c, s quark being massless ($m_u = m_d =
m_c = m_s = 0$) and the bottom-quark being massive. We use the
five-flavor scheme (5FS) in the leading-order (LO) and QCD NLO calculations. The
difference between adopting the five-flavor scheme and four-flavor
scheme is the ordering of the perturbative series for the
production cross section. In the four-flavor
scheme the perturbative series is
ordered strictly by powers of the strong coupling $\alpha_s$, while
in the 5FS the introduction of the b quark parton distribution
function (PDF) allows to resum terms of the form $\alpha_s^n
\ln(\mu^2/m_b^2)^m$ at all orders in $\alpha_s$. The calculations
are carried out in the 't Hooft-Feynman gauge. The {\sc FeynArts 3.4}
package\cite{fey} is adopted to generate Feynman diagrams and
convert them to the corresponding amplitudes. {\sc FormCalc 5.4}
programs\cite{formloop} are implemented to simplify the amplitudes.
As we know the cross section for the \gbWWb partonic process in the
SM should be the same as that for its charge conjugate subprocess
$g\bar b \to W^+W^-\bar b$, and the luminosity of the bottom-quark
in proton is the same as that of the antibottom quark. Therefore,
the production rates for both the $ W^+W^-b$ and the $W^+W^-\bar b$
productions at the LHC are identical. In the following sections we
present only the analytical calculations of the related partonic
process \gbWWb and the parent process \ppWWb unless otherwise
indicated.

\par
\subsection{Leading order cross sections }
\par
\begin{figure*}
\begin{center}
\includegraphics[scale=0.8]{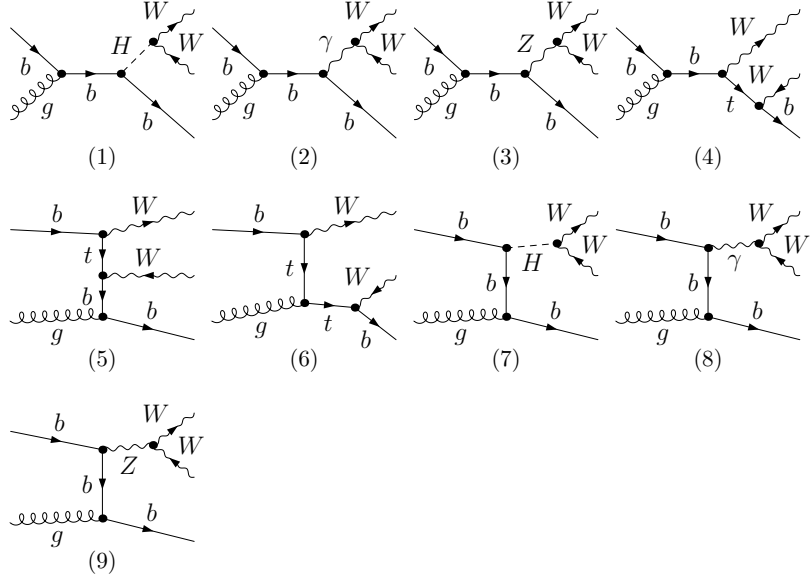}
\caption{ \label{fig1} The LO Feynman diagrams for the partonic
process \gbWWb, where the internal top-quark in Figs.1(4,6) and the
Higgs boson in Figs.1(1,7) may be on mass shell. }
\end{center}
\end{figure*}

We denote the concerned partonic process as $g(p_{1})+b(p_{2})\to
W^{+}(p_{3})+W^{-}(p_{4})+b(p_5)$. There are nine leading order (LO)
Feynman diagrams for this partonic process shown in Fig.\ref{fig1}.
There Figs.1(1-4) and Figs.1(5-9) are the s-channel and t-channel
diagrams for the partonic process, respectively. The LO cross
section for the partonic process \gbWWb is obtained by using the
following formula:
\begin{eqnarray}
\hat{\sigma}_{LO}(\hat{s}, gb \to W^+W^-b)= \frac{(2 \pi
)^4}{4|\vec{p}_1|\sqrt{\hat{s}}}\int \overline{\sum} |{\cal
M}_{LO}|^2 d\Phi_3,
\end{eqnarray}
where $d\Phi_3$ is the three-body phase-space element, and
$\vec{p}_1$ is the momentum of the initial gluon in the
center-of-mass system. The integration is performed over
the three-body phase space of the final particles $W^+W^-b$. The
summation is taken over the spins and colors of the initial and
final states, and the bar over the summation indicates averaging over
the intrinsic degrees of freedom of initial partons.

\par
In the LO calculations, the internal top quark and Higgs boson are
potentially resonant. We introduce the complex mass
scheme(CMS)\cite{CMS} to deal with the internal resonant top quark
and Higgs boson. In the CMS the complex masses of the unstable
top quark and Higgs boson should be taken everywhere in both
tree-level and one-loop level calculations. Then the gauge
invariance is conserved and the singularity poles of propagators for
real $p^2$ are avoided. The relevant complex masses are defined as
$\mu_t^2=m_t^2-im_t\Gamma_t$, $\mu_H^2=m_H^2-im_H\Gamma_H$, where
$m_{t}$ and $m_{H}$ are the conventional real masses, $\Gamma_{t}$
and $\Gamma_{H}$ represent the corresponding total widths of
the top quark and Higgs boson, and the poles of propagators are located
at $\mu_t^2$ and $\mu_H^2$ on the complex $p^2$ plane, respectively.
Since the unstable particles are involved in the loops for the
${\cal O}(\alpha_s)$ QCD corrections, we shall meet the calculations
of N point integrals with complex masses.

\par
The LO total cross section for $pp \to W^{+} W^{-}b+X$ can be
expressed as
\begin{eqnarray}\label{sigma_PP}
&& \sigma_{LO}(pp \to W^{+} W^{-}b+X)=   \nonumber \\
&& \int dx_A dx_B \left[ G_{g/A}(x_A,\mu_f)
G_{b/B}(x_B,\mu_f)\hat{\sigma}_{LO}(gb \to W^+W^-b,
x_{A}x_{B}s,\mu_f,\mu_r)+(A\leftrightarrow B)\right]. \nb \\
\end{eqnarray}
There $G_{i/P}$ $(i=g,b,~P=A,B)$ represent the PDFs of parton $i$
in proton $P$, $\mu_f$ and $\mu_r$
are the factorization and renormalization scales separately, and
$x_A$ and $x_B$ describe the momentum fractions of parton (gluon or
bottom-quark) in protons $A$ and $B$ respectively.

\par
\subsection{QCD NLO corrections }

\par
\subsubsection{General description} \label{Sec II.2.1}
\par
Our jet recombination procedure is based on the jet algorithm in
Ref.\cite{je_algo}. We consider the jet events with up to two
protojets involved in the final states. The candidate jet events
are applied by the jet recombination procedure with the following
criteria: If the two protojets $i$ and $j$ satisfy $\sqrt{\Delta
\eta^2 + \Delta \phi^2} < R \equiv 0.7$ (where $\Delta \eta$ and
$\Delta \phi$ are the differences of pseudorapidity and the azimuthal
angle between the two protojets), they are merged as one jet and
this jet event is called a one-jet event, whose four-momentum is
defined as $p_{ij,\mu}=p_{i,\mu}+p_{j,\mu}$. The transverse momentum
and rapidity of the merged jet are calculated from its four-momentum
by using the formulas $p_{ij,T}=\sqrt{p_{ij,x}^2+p_{ij,y}^2}$ and
$y_{ij}=\frac{1}{2}\ln{\left(\frac{E_{ij}+p_{ij,L}}{E_{ij}-p_{ij,L}}\right)}$.

\par
The QCD NLO correction to the $pp \to W^+W^-b+X$ process includes
the following components:
\begin{itemize}
 \item[(i)] The QCD one-loop virtual corrections to the partonic
       process $gb \to W^+W^-b$.
 \item[(ii)] The contribution of the real gluon emission partonic process
       $gb \to W^{+}W^{-}bg$.
 \item[(iii)] The contribution of the real light-(anti)quark emission partonic process
       $qb \to W^{+}W^{-}bq$, where $q$ denotes any of the light-(anti)quarks
       $u,d,c,s,\bar u,\bar d,\bar c,\bar s$.
 \item[(iv)] The corresponding contributions of the PDF counterterms.
 \item[(v)] The additional contributions of the PDFs for subtracting the quasicollinear
       mass singularity from massive bottom quark.
\end{itemize}

\par
In order to make a comparison, we use two schemes for the collection
of the finite contributions at ${\cal O}(\alpha_s^2 \alpha_{ew}^2)$.
In the QCD NLO contribution collection scheme-I, we collect only the five
components mentioned above for the QCD NLO corrections.
In the QCD NLO contribution collection scheme-II, in addition to the contributions
involved in the collection scheme-I, the QCD NLO corrections to the
$pp \to W^+W^-b + X$ process include the following three
additional contributions:
\begin{itemize}
 \item[(i)] The LO contribution of the (anti)bottom-quark radiative
       processes $q\bar{q}(gg) \to W^+W^-b\bar b$ ($q=u,d,c,s,b$) in the
       "merged one-jet phase space" (One half contribution belongs to
       the QCD NLO correction of the \ppWWb process, another half is to the
       correction of the \ppWWbbar process).
 \item[(ii)] The LO contributions of the $pp \to bb \to W^+W^-bb$ and
       $pp \to \bar b \bar b \to W^+W^-\bar b \bar b$
       processes in the "merged one-jet phase space" (The contribution from the
       former process belongs to the \ppWWb process, the latter part belongs
       to the \ppWWbbar process).
 \item[(iii)] The additional contributions of the PDFs and fragmentation functions (FFs)
       for subtracting the quasicollinear mass singularities in the
       $pp \to W^+W^-b\bar{b}(bb, \bar{b}\bar{b})$ processes.
\end{itemize}
The so-called "merged one-jet phase space" is defined where the
two final $b(\bar{b})$ jets are merged as one jet in the tagging
region of the detector, i.e., $\sqrt{\Delta \eta^2 + \Delta \phi^2}
< 0.7$, and the merged jet passes the constraints of $p_{T,j} >
25~GeV$ and $|y_{j}| < 2.5$. That is to say the two
(anti)bottom quarks cannot form two isolated hard jets, and
therefore we call the event a one-jet event. We notice that all the
contributions involved in the QCD NLO contribution collection scheme-II
are at ${\cal O}(\alpha_s^2 \alpha_{ew}^2)$.

\par
The dimensional regularization method in $D=4-2 \epsilon$ dimensions
is used to isolate the UV and IR singularities. We split each
collinear counterterm of the PDF, $\delta G_{i/P}(x,\mu_f)$ ($P=$
proton, $i=g,u,\bar{u}$,$d,\bar{d}$,$c,\bar{c}$,$s,\bar{s}$), into two
parts: the collinear gluon emission part $\delta
G_{i/P}^{(gluon)}(x,\mu_f)$ and the collinear light-quark emission
part $\delta G_{i/P}^{(quark)}(x,\mu_f)$. The analytical expressions
are presented as follows.
\begin{eqnarray}\label{PDFcounterterm1}
&& \delta G_{q(g)/P}(x,\mu_f) = \delta G_{q(g)/P}^{(gluon)}(x,\mu_f)
                  +\delta G_{q(g)/P}^{(quark)}(x,\mu_f),
                ~~(q = u, \bar{u}, d, \bar{d}, c, \bar{c}, s, \bar{s} ),
\end{eqnarray}
where
\begin{eqnarray}\label{PDFcounterterm2}
&& \delta G_{q(g)/P}^{(gluon)}(x,\mu_f) =
   \frac{1}{\epsilon} \left[
                      \frac{\alpha_s}{2 \pi}
                      \frac{\Gamma(1 - \epsilon)}{\Gamma(1 - 2 \epsilon)}
                      \left( \frac{4 \pi \mu_r^2}{\mu_f^2} \right)^{\epsilon}
                      \right]
   \int_x^1 \frac{dz}{z} P_{qq(gg)}(z) G_{q(g)/P}(x/z,\mu_f), \nonumber \\
&& \delta G_{q/P}^{(quark)}(x,\mu_f) =
   \frac{1}{\epsilon} \left[
                      \frac{\alpha_s}{2 \pi}
                      \frac{\Gamma(1 - \epsilon)}{\Gamma(1 - 2 \epsilon)}
                      \left( \frac{4 \pi \mu_r^2}{\mu_f^2} \right)^{\epsilon}
                      \right]
   \int_x^1 \frac{dz}{z} P_{qg}(z) G_{g/P}(x/z,\mu_f),  \nonumber \\
&& \delta G_{g/P}^{(quark)}(x,\mu_f) =
   \frac{1}{\epsilon} \left[
                      \frac{\alpha_s}{2 \pi}
                      \frac{\Gamma(1 - \epsilon)}{\Gamma(1 - 2 \epsilon)}
                      \left( \frac{4 \pi \mu_r^2}{\mu_f^2} \right)^{\epsilon}
                      \right]
   \sum_{q=u,\bar{u}}^{d,\bar{d}, c, \bar {c}, s, \bar {s}}
   \int_x^1 \frac{dz}{z} P_{gq}(z) G_{q/P}(x/z,\mu_f),
\end{eqnarray}
and the explicit expressions for the splitting functions
$P_{ij}(z),~(ij=qq,qg,gq,gg)$ can be found in Ref.\cite{19}.

\par
\subsubsection{Virtual and real emission corrections to \gbWWb}
\par
\begin{figure*}
\begin{center}
\includegraphics[scale=0.8]{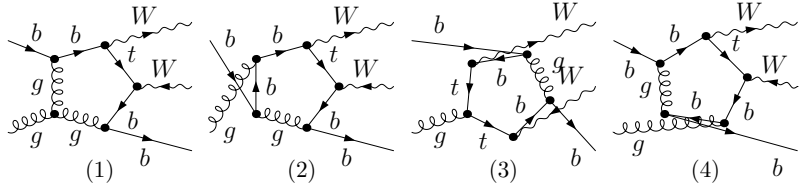}
\caption{ \label{fig2} The pentagon Fynman diagrams for the partonic
process $gb \to W^+W^-b$. }
\end{center}
\end{figure*}

The amplitude at the one-loop level for the partonic process \gbWWb in the SM
contains the contributions of the self-energy, vertex, box,
pentagon and counterterm graphs. In Fig.\ref{fig2} the four pentagon
Feynman diagrams are shown as representatives.

\par
In order to remove the UV divergences in virtual corrections, we
need to renormalize the strong coupling constant, the masses and
wave functions of the relevant colored particles. In our
calculations we introduce the following renormalization constants:
\begin{eqnarray}
\label{defination of renormalization constants} \psi_{b(t)}^{0,L,R}
&=& \left(1+\frac{1}{2}\delta Z_{b(t)}^{L,R}\right)
\psi_{b(t)}^{L,R},~~~~~m^{0}_{b} = m_{b}+\delta m_{b},~~ \nb \\
\mu^{0}_{t} &=& \mu_{t}+\delta \mu_{t},~~~~G_{\mu}^0 = (1+
\frac{1}{2}\delta Z_g)G_{\mu},~~~~g_s^0 = g_s+\delta g_s,
\end{eqnarray}
where $g_s$ denotes the strong coupling constant, $m_b$ and $\mu_t$
are bottom-quark mass and top-quark complex mass, respectively, and
$\psi_b^{L,R},~\psi_t^{L,R}$, and $G_{\mu}$ denote the fields of
bottom quark, top quark and gluon, separately. The masses and wave
functions of the colored fields are renormalized
in the on-shell scheme, and the relevant counterterms are
expressed as
\begin{eqnarray} \label{CT-q}
\delta Z_{b}^{L,R}&=&- \frac{\alpha_s(\mu_r)}{3 \pi}
\left[\Delta_{UV}+2\Delta_{IR}+4+3\ln\left(\frac{\mu_r^2}{m_{b}^2}\right)\right], \nb \\
\delta Z_{t}^{L,R}&=&- \frac{\alpha_s(\mu_r)}{3 \pi}
\left[\Delta_{UV}+2\Delta_{IR}+4+3\ln\left(\frac{\mu_r^2}{\mu_{t}^2}\right)\right], \nb \\
\frac{\delta m_{b}}{m_{b}} &=& - \frac{\alpha_s(\mu_r)}{3 \pi}
\left\{ 3\left[ \Delta_{UV} + \ln\left( \frac{\mu_r^2}{m_{b}^2}
\right) \right]+4
\right \}, \nb \\
\frac{\delta \mu_{t}}{\mu_{t}} &=& - \frac{\alpha_s(\mu_r)}{3 \pi}
\left\{ 3\left[ \Delta_{UV} + \ln\left( \frac{\mu_r^2}{\mu_{t}^2}
\right) \right]+4
\right \}, \nb \\
\delta Z_g &=&- \frac{\alpha_s(\mu_r)}{2
\pi}\left\{-\frac{1}{2}\Delta_{UV}+\frac{7}{6}\Delta_{IR}+
\frac{1}{3}\left[\ln\left(\frac{\mu_r^2}{m_b^2}\right)+
\ln\left(\frac{\mu_r^2}{\mu_{t}^2}\right)\right]\right\},
\end{eqnarray}
where$\Delta_{UV}=1/\epsilon_{UV} -\gamma_E +\ln(4\pi)$ and
$\Delta_{IR}=1/\epsilon_{IR} -\gamma_E +\ln(4\pi)$.

\par
For the renormalization of the strong coupling constant $g_{s}$, we
adopt the $\overline{MS}$ scheme at the renormalization scale $\mu_{r}$,
except that the divergences associated with the massive top- and
bottom-quark loops are subtracted at zero momentum\cite{gs}. Then
the counterterm of the strong coupling constant $g_{s}$ can be
obtained as
\begin{eqnarray} \label{CT-g}
&& \frac{\delta g_s}{g_s}= -\frac{\alpha_s(\mu_r)}{4\pi}
\left[\frac{\beta_0}{2}\Delta_{UV} +\frac{1}{3}\ln\frac{m_{b}^2}
{\mu_{r}^2}+\frac{1}{3}\ln\frac{\mu_{t}^2} {\mu_{r}^2}\right],
\end{eqnarray}
where $\beta_0=\frac{11}{3}N-\frac{2}{3}N_{f}-\frac{4}{3}$, and the
numbers of colors and active flavors are taken as $N=3$ and $N_f=4$
respectively. As shown in Eqs.(\ref{CT-q}) and (\ref{CT-g}), the
renormalizaion constants $\delta Z_g$, $\delta Z_b^{L,R}$, $\delta
m_b$ and $\delta g_s/g_s$ contain the terms of $\alpha_s \ln \left(
\frac{\mu_r^2}{m_b^2} \right)$, which is divergent when $m_b \to 0$.
Therefore, the amplitude of the counterterms for the subprocess $gb
\to W^+W^-b$ also contains the quasicollinear mass-singular term.
By adding the virtual and real corrections to the contributions of
counterterms, the ${\cal O}(\alpha_s^2 \alpha_{ew}^2)$ NLO QCD
corrected partonic cross section still involves the quasicollinear
mass-singular term, which could violate the convergence of the
perturbative series. But this quasicollinear mass-singular term can
be canceled exactly by the additional contributions of the PDFs and
the FFs for subtracting the quasicollinear mass singularity due to
the massive bottom quark, as explained in the following subsection.

\par
Because we use the CMS to deal with the possible top-quark and Higgs-boson 
resonances, the normal one-loop integrals must be continued
onto the complex plane. The formulas for calculating the
IR-divergent integrals with complex internal masses in the dimensional
regularization scheme are obtained by analytically continuing the
expressions in Ref.\cite{Stefan} onto the complex plane. The
numerical evaluations of IR-safe N point ($n=1,2,3,4,5$) integrals
with complex masses, are implemented by using the expressions
analytically continued onto the complex plane from those presented in
Refs.\cite{OneTwoThree,Four,Five}. In this way, we created our
in-house subroutines to isolate analytically the IR singularities in
integrals and calculate numerically one-loop integrals with complex
masses based on the {\sc LoopTools-2.4} package\cite{formloop}\cite{ff}.

\par
The corrections due to the real gluon and (anti)light-quark emission
partonic processes, $b(p_1)~g[q,\bar q](p_2)\to
W^{+}(p_3)~W^-(p_4)~b(p_5)~g[q,\bar q](p_6)$$(q=u,d,c,s)$, are dealt
with by using the two cutoff phase-space slicing
method\cite{19}. The soft IR singularity of the real gluon emission
subprocess can be isolated by introducing an arbitrary soft cutoff
$\delta_{s}$, to separate the $2 \to 4$ phase space into two
regions, $E_{6}\leq\delta_{s}\sqrt{\hat{s}}/2$ (soft gluon region)
and $E_{6}>\delta_{s}\sqrt{\hat{s}}/2$ (hard gluon region). The
cutoff $\delta_{c}$ decomposes the real hard gluon/(anti)light-quark
emission phase-space region into hard collinear ($HC$) region,
$\hat{s}_{26} < \delta_c \hat{s}$ (where
$\hat{s}_{ij}=(p_{i}+p_{j})^{2}$), and noncollinear
($\overline{HC}$) region, $\hat{s}_{26} > \delta_c \hat{s}$, in
order to isolate the collinear singularity from the IR-safe
region. The integration over the $\overline{HC}$ region of phase
space can be performed in the four-dimensions by using the program
based on the VEGAS, a Monte Carlo integrator\cite{guolei}. Then the
cross section for the real emission processes can be written as
\begin{equation}
\label{sigmaR} \Delta \sigma_{R}=\Delta \sigma_{S}+\Delta \sigma_{H}
=\Delta \sigma_{S}+\Delta \sigma_{HC}+\Delta \sigma_{\overline{HC}}.
\end{equation}

\par
The UV singularities in the virtual corrections are canceled by the
contributions of all the related counterterms. Soft and collinear IR
singularities are also involved in virtual corrections. After
combining the contributions of the real gluon/light-quark emission
processes and the PDF counterterms $\delta G_{q(g)/P}$ with the
virtual contributions together, these IR singularities are exactly
vanished. These cancelations have been verified numerically in our
numerical calculations.

\vskip 5mm
\subsubsection{Subtraction of quasicollinear mass singularity}\label{quasi_coll}
\par
Since the bottom-quark mass is retained within the whole
calculation, the PDFs of (anti)bottom quark and gluon (up to the
$\alpha_s$ order) contain the large logarithm terms of $\left(
\alpha_s\ln \frac{\mu_f^2}{m_b^2} \right)$. The large logarithm terms in
the $G_{b(\bar{b})/P}(x, \mu_f)$ and $G_{g/P}(x, \mu_f)$ PDFs arise
from the (anti)bottom-quark emission off the (anti)bottom quark or
gluon and gluon emission off the (anti)bottom quark, respectively.
They are just the QCD NLO counterterms of $G_{b(\bar{b})/P}(x,
\mu_f)$ and $G_{g/P}(x, \mu_f)$ which are finite with a nonzero
bottom-quark mass. The isolation of these quasicollinear
mass-singular terms are similar to that in the conventional massless
parton model approach.

\par
Analogous to the PDFs of light quarks ($q = u, \bar{u}, d, \bar{d},
c, \bar{c}, s, \bar{s}$) in Eq.(\ref{PDFcounterterm1}), the QCD NLO
counterterm of $G_{b(\bar{b})/P}(x, \mu_f)$ can be split into the gluon
emission part and the $\bar{b}(b)$ emission part.
\begin{eqnarray}
\delta G_{b(\bar{b})/P}(x, \mu_f) = \delta
G_{b(\bar{b})/P}^{(gluon)}(x, \mu_f) + \delta
G_{b(\bar{b})/P}^{(\bar{b}(b))}(x, \mu_f).
\end{eqnarray}
The explicit expressions for these two components can be expressed
as\cite{ini_coll,FFS}
\begin{eqnarray}
&& \delta G_{b(\bar{b})/P}^{(gluon)}(x, \mu_f) =
   -\frac{\alpha_s}{2 \pi}
   \int_{x}^1\frac{d z}{z}\left[ P_{bb(\bar{b}\bar{b})}(z)
   \left(\ln \frac{\mu_f^2}{m_b^2} -2 \ln(1-z) -1 \right) \right]_{+}
   G_{b(\bar{b})/P}(x/z,\mu_f)  \nonumber \\
&& \delta G_{b(\bar{b})/P}^{(\bar{b}(b))}(x, \mu_f) =
   -\frac{\alpha_s}{2 \pi}
   \int_{x}^1\frac{d z}{z} P_{bg(\bar{b}g)}(z)
   \left( \ln \frac{\mu_f^2}{m_b^2} \right)
   G_{g/P}(x/z,\mu_f),
\end{eqnarray}
where the $[\dots]_+$ prescription is defined by
\begin{eqnarray}
 \int_x^1 dz \left[ f(z) \right]_+ g(z)
= \int_x^1 dz f(z) g(z)- \int_0^1 dz f(z) g(1)~~.
\end{eqnarray}
Obviously these two counterterms are proportional to $\left(
\alpha_s\ln \frac{\mu_f^2}{m_b^2} \right)$ and finite with nonzero
bottom-quark mass.

\par
The QCD NLO counterterm of $G_{g/P}(x, \mu_f)$ is also divided into
two components. Among them one is the divergent, whose analytical
expression is proportional to $\frac{1}{\epsilon} \left[
\frac{\alpha_s}{2 \pi} \frac{\Gamma(1 - \epsilon)}{\Gamma(1 - 2
\epsilon)} \left( \frac{4 \pi \mu_r^2}{\mu_f^2} \right)^{\epsilon}
\right]$, and has been given in Eqs.(\ref{PDFcounterterm1}) and
(\ref{PDFcounterterm2}). Another component is the additional part,
$\delta G_{g/P}^{(add)}(x, \mu_f)$, which is proportional to $\left(
\alpha_s\ln \frac{\mu_f^2}{m_b^2} \right)$, i.e.,
\begin{eqnarray}
 \delta G_{g/P}^{(add)}(x, \mu_f)
 &=&
 \delta G_{g/P}^{(b)}(x, \mu_f) + \delta G_{g/P}^{(\bar{b})}(x, \mu_f)
 \nonumber \\
 &=& -\frac{\alpha_s}{2 \pi}
   \int_{x}^1\frac{d z}{z} P_{gb}(z)
   \left( \ln \frac{\mu_f^2}{m_b^2} \right)
   G_{b/P}(x/z,\mu_f) \nonumber \\
 && -\frac{\alpha_s}{2 \pi}
   \int_{x}^1\frac{d z}{z} P_{g\bar{b}}(z)
   \left( \ln \frac{\mu_f^2}{m_b^2} \right)
   G_{\bar{b}/P}(x/z,\mu_f).
\end{eqnarray}

\par
In order to include the contributions of these quasicollinear
mass-singular counterterms of the PDFs in the total NLO cross
section, we replace the PDFs in the expressions for $\sigma_{LO}(pp
\to W^+W^-b + X)$ and $\sigma_{LO}(pp \to W^+W^-\bar{b} + X)$ (see
Eq.(\ref{sigma_PP}) for $\sigma_{LO}(pp \to W^+W^-b + X)$ and
similar one for $\sigma_{LO}(pp \to W^+W^-\bar{b} + X)$) as
\begin{eqnarray}
  && G_{b(\bar{b})/P}(x, \mu_f) \longrightarrow  G_{b(\bar{b})/P}(x, \mu_f)
  + \delta G_{b(\bar{b})/P}^{(gluon)}(x,
  \mu_f) + \delta G_{b(\bar{b})/P}^{(\bar{b}(b))}(x, \mu_f), \nb \\
  && G_{g/P}(x, \mu_f)  \longrightarrow  G_{g/P}(x, \mu_f) +
  \delta G^{(gluon)}_{g/P} (x, \mu_f) + \delta G^{(quark)}_{g/P} (x, \mu_f)+
  \delta G_{g/P}^{(add)}(x, \mu_f). \nb \\
\end{eqnarray}
Then we obtain the following five additional contributions of the
PDFs:
\begin{eqnarray}\label{additional CT}
\textrm{(i)} ~~&&\sigma_{PDF}(W^+W^-bg) \nonumber \\
  &=&
  \int dx_A dx_B
  \left[
  G_{g/A}(x_A, \mu_f) \delta G_{b/B}^{(gluon)}(x_B, \mu_f)
  \hat{\sigma}_{LO}(g b \to W^+ W^- b) + (A \leftrightarrow B)
  \right], \nonumber \\
\textrm{(ii)} ~~&&\sigma_{PDF}(W^+W^-\bar{b}g) \nonumber \\
  &=&
  \int dx_A dx_B
  \left[
  G_{g/A}(x_A, \mu_f) \delta G_{\bar{b}/B}^{(gluon)}(x_B, \mu_f)
  \hat{\sigma}_{LO}(g \bar{b} \to W^+ W^- \bar{b}) + (A \leftrightarrow B)
  \right], \nonumber \\
\textrm{(iii)} ~~&&\sigma_{PDF}(W^+W^-b b) \nonumber \\
  &=&
  \int dx_A dx_B
  \left[
  G_{b/A}(x_A, \mu_f) \delta G_{g/B}^{(b)}(x_B, \mu_f)
  \hat{\sigma}_{LO}(g b \to W^+ W^- b) + (A \leftrightarrow B)
  \right], \nonumber \\
\textrm{(iv)} ~~&&\sigma_{PDF}(W^+W^-\bar{b}\bar{b}) \nonumber \\
  &=&
  \int dx_A dx_B
  \left[
  G_{\bar{b}/A}(x_A, \mu_f) \delta G_{g/B}^{(\bar{b})}(x_B, \mu_f)
  \hat{\sigma}_{LO}(g \bar{b} \to W^+ W^- \bar{b}) + (A \leftrightarrow B)
  \right], \nonumber \\
\textrm{(v)} ~~&&\sigma_{PDF}(W^+W^-b\bar{b}) \nonumber \\
  &=&
  \sigma_{PDF}^{(gg)}(W^+W^-b\bar{b}) ~+~ \sigma_{PDF}^{(b\bar{b})}(W^+W^-b\bar{b}) \nonumber \\
  &=&
  \int dx_A dx_B
  \left[ G_{g/A}(x_A, \mu_f) \delta G_{b/B}^{(\bar{b})}(x_B, \mu_f) \hat{\sigma}_{LO}(g b \to W^+ W^- b)
  \right.
  \nonumber \\
  &&~~~~~~~~~~~~+
  \left.
  G_{g/A}(x_A, \mu_f) \delta G_{\bar{b}/B}^{(b)}(x_B, \mu_f)
  \hat{\sigma}_{LO}(g \bar{b} \to W^+ W^- \bar{b})
  +
  (A \leftrightarrow B) \right]
  \nonumber \\
  &+&
  \int dx_A dx_B
  \left[ G_{b/A}(x_A, \mu_f) \delta G_{g/B}^{(\bar{b})}(x_B, \mu_f) \hat{\sigma}_{LO}(g b \to W^+ W^- b)
  \right.
  \nonumber \\
  &&~~~~~~~~~~~~+
  \left.
  G_{\bar{b}/A}(x_A, \mu_f) \delta G_{g/B}^{(b)}(x_B, \mu_f)
  \hat{\sigma}_{LO}(g \bar{b} \to W^+ W^- \bar{b})
  +
  (A \leftrightarrow B) \right].
\end{eqnarray}

\par
From the above discussion we can see that the large logarithm in the
cross section $\sigma(pp \to gg \to W^+W^-b\bar b+X)$ is canceled
exactly by that in the additional PDF contribution
$\sigma_{PDF}^{(gg)}(W^+W^-b\bar b)$ shown in Eq.(\ref{additional
CT}). To cancel the quasicollinear mass singularity of the $pp \to
b\bar{b} \to W^+W^- b\bar{b} + X$ process, both the additional
contributions of the PDFs and the FFs should be included, since the
quasicollinear mass singularity arises not only from the
(anti)bottom-quark emission off the initial (anti)bottom quark, but
also the gluon splitting to the $b\bar{b}$ pair in the final states. For
the $pp \to q \bar q \to W^+ W^- b \bar b +X$ ($q=u,d,c,s$)
processes there also exists a large logarithm corresponding to the
final gluon splitting to the $b\bar{b}$ pair, which can be absorbed by
the additional FF contribution. The FF additional contribution to
the cross section can be expressed as\cite{FFS}
\begin{eqnarray}\label{FFSeq}
 & & \sigma_{FF}^{(q \bar{q})} (W^+ W^- b \bar{b}) \nonumber \\
 &=& -\int dx_A dx_B dz \Big [G_{q/A}(x_A,\mu_f)G_{\bar{q}/B}(x_B,\mu_f)
 \hat{\sigma}_{LO}(q \bar{q} \to W^+ W^- g)\frac{\alpha_s}{2\pi}
 P_{bg}(z)\left (\ln \frac{\mu^2_f}{m_b^2} \right ) \nonumber \\
 & & + (A \leftrightarrow B)\Big ],~~(q=u,d,c,s,b).
\end{eqnarray}

\vskip 5mm
\subsection{Cancellation of quasicollinear mass-singularity}
\par
The quasicollinear mass singularity of the real gluon emission
process $pp \to g b \to W^+ W^- b g + X$ arises from the gluon
emission off either the initial or the final bottom quark. The large
logarithm term corresponding to the gluon emission off the initial
bottom quark is canceled by that of $\sigma_{PDF}(W^+W^-bg)$, while
the large logarithm term corresponding to the gluon emission off the
final bottom quark is canceled by that of the virtual corrections to
the $pp \to W^+ W^- b + X$ process exactly. For the real light-quark
emission processes $pp \to qb \to W^+ W^- bq + X$ ($q = u, \bar{u},
d, \bar{d}, c, \bar{c}, s, \bar{s}$), the final bottom quark is
noncollinear to the initial bottom quark since a bottom jet should
be detected in the final states. Therefore, the real light-quark
emission processes $pp \to qb \to W^+ W^- bq + X$ ($q = u, \bar{u},
d, \bar{d}, c, \bar{c}, s, \bar{s}$) do not contain quasicollinear
mass singularity.

\par
After the cancelation of quasicollinear mass singularity explained
above, we get the total QCD NLO correction to the $pp \to W^+ W^- b
+ X$ process by adopting the QCD NLO contribution collection
scheme-I (see next section), which is free of the contribution from
the large logarithm term $\left( \alpha_s\ln \frac{\mu_f^2}{m_b^2}
\right)$, expressed as
\begin{eqnarray}
 && \Delta \sigma_{NLO}(pp \to W^+W^-b + X) \nonumber \\
 &=&
 \sum_{q=u,d,c,s}^{\bar{u},\bar{d},\bar{c},\bar{s}}
 \Delta \sigma_{R}(pp \to q b \to W^+W^- b q +X)+
 \left[ \Delta \sigma_{V}(pp \to g b \to W^+W^-b + X) \right. \nonumber \\
 &+&\left. \Delta \sigma_{R}(pp \to g b \to W^+W^- b g + X) +
 \sigma_{PDF}(W^+W^-b g) \right],
\end{eqnarray}
We find that $ \Delta \sigma_{NLO}(pp \to W^+W^-b + X)$ is
convergent when $m_b \to 0$. Analogously, the total QCD NLO
correction to the $pp \to W^+ W^- \bar{b} + X$ process, which is the
summation of the virtual, real corrections and
$\sigma_{PDF}(W^+W^-\bar{b}g)$, is also finite when $m_b \to 0$.

\par
There are 28 Feynman diagrams for the $pp \to b b \to W^+ W^- b b +
X$ process at the ${\cal O}(\alpha_s^2 \alpha_{ew}^2)$. We can find
that one (and only one) of the two final bottom quarks is emitted
off the initial bottom quarks for each of these Feynman diagrams,
therefore, the cross section for the $pp \to b b \to W^+ W^- b b +
X$ process contains the large logarithm term of $\alpha_s \ln \left(
\frac{\mu_r^2}{m_b^2} \right)$. This quasicollinear mass singular
term is canceled exactly by that of $\sigma_{PDF}(W^+W^-b b)$.
Similarly, the large logarithm terms in $\sigma(pp \to \bar{b}
\bar{b} \to W^+ W^- \bar{b} \bar{b} + X)$ and
$\sigma_{PDF}(W^+W^-\bar{b} \bar{b})$ are canceled by each other.
Then the total cross sections for the $pp \to W^+ W^- b b + X$ and
$pp \to W^+ W^- \bar{b} \bar{b} + X$ processes defined as
\begin{eqnarray}
  \sigma_{tot}(pp \to W^+ W^- b b + X)
  &=&
  \sigma(pp \to b b \to W^+ W^- b b + X)
  + \sigma_{PDF}(W^+W^-b b) \nonumber \\
  \sigma_{tot}(pp \to W^+ W^- \bar{b} \bar{b} + X)
  &=&
  \sigma(pp \to \bar{b} \bar{b} \to W^+ W^- \bar{b} \bar{b} + X)
  + \sigma_{PDF}(W^+W^- \bar{b} \bar{b}),
\end{eqnarray}
are convergent when $m_b \to 0$.

\par
The quasicollinear mass singularity sources in the $pp \to W^+ W^-
b \bar{b} + X$ process are listed as follows:
\begin{itemize}
\item[(i)] $pp \to q\bar{q} \to W^+W^- b\bar{b} + X~(q=u,d,c,s)$:~~the internal gluon
splitting into $b\bar{b}$-pair.
\item[(ii)] $pp \to gg \to W^+W^- b\bar{b} + X$:~~the (anti)bottom quark emitting off the initial
gluon.
\item[(iii)] $pp \to b\bar{b} \to W^+W^- b\bar{b} + X$:~~the (anti)bottom-quark
emitting off the initial (anti)bottom quark and the internal gluon
splitting into $b\bar{b}$ pair.
\end{itemize}
The LO total cross section for the $pp \to W^+ W^- b \bar{b} + X$
process defined as
\begin{eqnarray}\label{wwbbbar}
  &&\sigma_{tot}(pp \to W^+ W^- b \bar{b} + X) = \nb \\
  &&\sum_{q = u, d, c, s}\left[ \sigma(pp \to q \bar{q} \to W^+ W^- b \bar{b} + X)
  + \sigma_{FF}^{(q \bar{q})}(W^+ W^- b \bar b)\right] \nonumber \\
   &+& \Big[ \sigma(pp \to b \bar{b} \to W^+ W^- b \bar{b} + X)
  + \sigma_{PDF}^{(b\bar{b})}(W^+W^-b\bar{b})+\sigma_{FF}^{(b \bar{b})}(W^+ W^- b \bar b) \Big] \nb \\
  &+& \Big[ \sigma(pp \to g g \to W^+ W^- b \bar{b} + X)
  + \sigma_{PDF}^{(gg)}(W^+W^-b\bar{b}) \Big],
\end{eqnarray}
which is finite when $m_b \to 0$ and would be involved in the QCD
NLO corrections to the $pp \to W^+W^- b + X$ and $pp \to W^+W^-
\bar{b} + X$ processes.

\vskip 5mm
\section{Numerical results and discussions }\label{numres}
\par
\subsection{Input parameters}\label{parameters}
\par
In this work we take one-loop and two-loop running $\alpha_{s}$ in
the LO and NLO calculations, respectively\cite{hepdata}. For
simplicity we set the factorization scale and the renormalization
scale being equal (i.e., $\mu=\mu_f=\mu_r$) and take
$\mu=\mu_0=m_b/2+m_W$ in default unless otherwise stated. Throughout
this paper, we take
$\alpha_{ew}(m_Z^2)^{-1}|_{\overline{MS}}=127.925$,
$m_W=80.398~GeV$, $m_Z=91.1876~GeV$,
$\sin^2\theta_w=1-\left(\frac{m_W}{m_Z}\right)^2=0.222646$, and set
quark masses as $m_u=m_d=m_c=m_s=0$, $m_b=4.2~GeV$ and
$m_t=171.2~GeV$\cite{hepdata}. The colliding energy in the
proton-proton center-of-mass system is taken as $\sqrt s=14~TeV$ for
the LHC. The CKM matrix elements are set as
\begin{eqnarray}\label{CKM}
 V_{CKM} &=& \left(
\begin{array}{ccc}
    V_{ud} \ &  V_{us} \ &  V_{ub} \\
    V_{cd} \ &  V_{cs} \ &  V_{cb} \\
    V_{td} \ &  V_{ts} \ &  V_{tb} \\
\end{array}
    \right)=\left(
\begin{array}{ccc}
     0.97418 \ &  0.22577 \ &  0 \\
    -0.22577 \ &  0.97418 \ &  0 \\
       0 \ &  0 \ &  1 \\
\end{array}  \right).
\end{eqnarray}
We adopt the CTEQ6L1 and CTEQ6M PDFs in the LO and NLO calculations,
respectively. The QCD parameter $\Lambda_4^{LO}=215~MeV$ for the
CTEQ6L1 at the LO, and $\Lambda_4^{\overline{MS}}=326~MeV$ for the
CTEQ6M at the NLO\cite{cteq}.

\par
Since we take $V_{tb}\sim 1$, the decay of the top quark is dominated by
the $t \to W^+ b$ decay mode, and the total decay width of the top quark
is approximately equal to the decay width of $t \to W^+b$.
Neglecting the terms of order $m_b^2/m_t^2$, $\alpha_s^2$ and
$(\alpha_s/\pi)m_W^2/m_t^2$, the width predicted in the SM is
\cite{twidth}
\begin{equation}
\Gamma_t=\frac{\alpha_{ew}
m_t^3}{16m_W^2s_W^2}\left(1-\frac{m_W^2}{m_t^2}\right)^2
\left(1+\frac{2m_W^2}{m_t^2}\right) \left[
1-\frac{2\alpha_s}{3\pi}\left(\frac{2\pi^2}{3}-\frac{5}{2}
\right)\right].
\end{equation}
By taking
$\alpha_{ew}=\alpha_{ew}(m_Z^2)|_{\overline{MS}}=1/127.925$ and
$\alpha_s(m_t^2)=0.1024$, we obtain $\Gamma_t=1.3692~GeV$. The
reasonable physical decay width of the Higgs-boson is obtained by
employing the program Hdecay\cite{hdecay}, where the partial decay
width $\Gamma (H^0 \to q\bar q)$ is calculated up to the ${\cal
O}(\alpha_s^3 \alpha_{ew})$. Then we obtain $\Gamma_H = 0.2965
\times 10^{-2}$, $1.704 \times 10^{-2}$ and $0.6511$ $GeV$ for $m_H
= 120$, $150$ and $180~GeV$, respectively.

\par
The verification of the independence of the total QCD NLO correction
to the \ppWWb process on the two cutoffs, $\delta_s$ and $\delta_c$,
is made. It shows that the total QCD NLO correction
$\Delta\sigma_{NLO}$, which is the summation of the three-body and
four-body cross sections, is independent of the two cutoffs within
the statistical errors. That independence of the full QCD NLO
correction to the \ppWWb process on the cutoffs $\delta_s$ and
$\delta_c$ provides an indirect check for the correctness of the
calculations. In the further numerical calculations, we fix
$\delta_s=1 \times 10^{-3}$ and $\delta_c=\delta_s/50$.

\vskip 5mm
\subsection{Event selection criteria }
\par
In this subsection we present the description of the event selection
criteria. After applying the jet recombination procedure to the
protojet events of the $pp \to W^+W^-b(\bar{b})+X$ up to the QCD
NLO, we obtain the one-jet events and two-jet events. The one-jet
event contains only a $b(\bar{b})$ jet or a merged jet involving at
least one $b(\bar{b})$ quark, while the two-jet event contains two
isolated hard jets which are $b(\bar{b})$ jet and
gluon/light-quark jet, respectively. In the following we present
detailed criteria for how to treat the one-jet and two-jet events.

\par
(1) For the one-jet events, we collect the events with the
constraints on the jet as $p_{T,j}>25~GeV~$ and $|y_{j}|<2.5$.

\par
(2) For the two-jet events (i.e., $W^+ W^- b(\bar{b}) j$ ($j = g, q~
{\rm and}~q = u, \bar{u}, d, \bar{d}, c, \bar{c}, s, \bar{s}$)
production), we treat the hard gluon/light-quark jet ($j$ jet),
which is noncollinear to the $b(\bar{b})$ jet, either inclusively
or exclusively. The inclusive and exclusive two-jet event
selection schemes are declared as follows:
\begin{itemize}
\item[(i)] In the inclusive two-jet event selection scheme both one- and
two-jet events are included, no further restriction is applied on
the $j$ jet except for the $b(\bar{b})$ jet. The $b(\bar{b})$ jet
should satisfy $p_{T,b(\bar{b})} > 25~GeV$ and $|y_{b(\bar{b})}| <
2.5$.

\item[(ii)] In the exclusive two-jet event selection scheme,
the one-jet events are accepted while the two-jet events are
rejected. Therefore, besides the $W^+W^-b(\bar{b})$ production
events, we accept the $W^+W^-b(\bar{b})j$ production events only
when the $b(\bar{b})$ jet satisfies $p_{T,{b(\bar{b})}} > 25~GeV$
and $|y_{b(\bar{b})}| < 2.5$, and another gluon/light-quark jet
($j$ jet) passes the constraint of either $p_{T,j} < 25~GeV$ or
$|y_j|> 2.5$. That means the second $j$ jet is undetectable or there
is no second separable jet being observed for the $W+ W^- b(\bar{b})
j$ production events.
\end{itemize}

\par
The cross sections for the $pp \to W^+W^-b\bar b+X$, $pp \to
W^+W^-bb+X$ and $pp \to W^+W^-\bar{b}\bar{b}+X$ processes combined
with the corresponding additional contributions of the PDFs and the
FFs shown in Eqs.(\ref{additional CT})and (\ref{FFSeq}), are free of
the quasicollinear mass singularity induced by the $\alpha_s \ln
\left( \mu_{f,r}^2/m_b^2 \right)$ term. Part of these contributions
are regarded as the QCD NLO corrections to the \ppWWbb process in
the QCD NLO contribution collection scheme-II, where the two
$b(\bar{b})$ protojets are merged as one jet. In Table \ref{tab2},
we present the numerical results of the integrated cross sections
for the $pp \to W^+ W^- b \bar{b}+X$, $W^+ W^- b b+X$, $W^+ W^-
\bar{b} \bar{b}+X$ processes over three different regions of final
state phase space with $m_H = 120$, $150$ and $180~GeV$,
respectively.

\vskip 5mm
\subsection{Dependence on energy scale }
\par
\begin{figure}
\begin{center}
\includegraphics[scale=1.0]{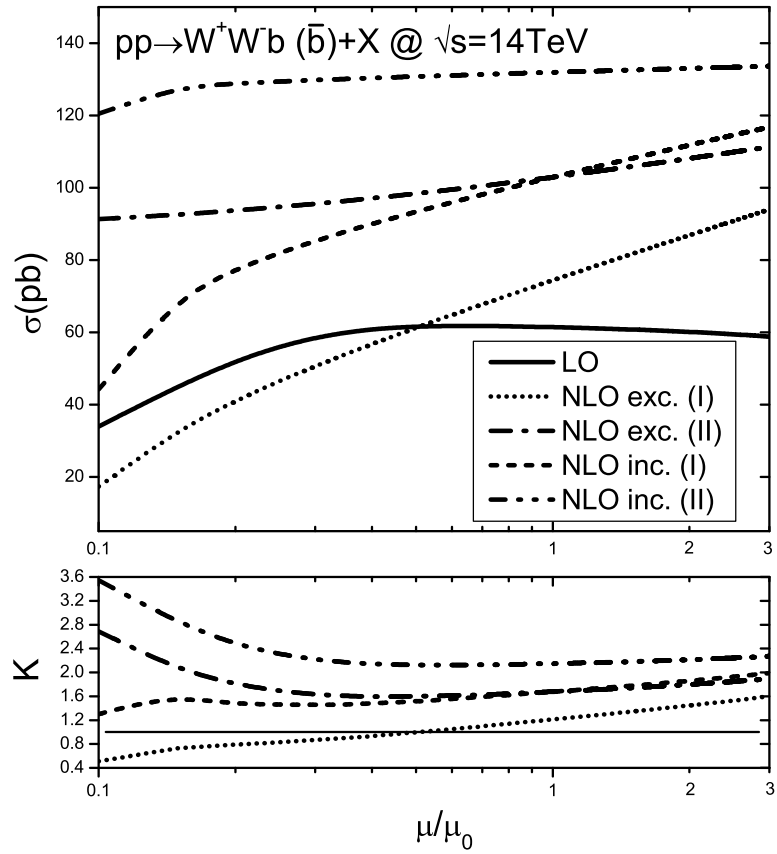}
\caption{\label{fig3} The dependence of the LO and NLO QCD corrected
total cross sections for the \ppWWb and \ppWWbbar processes on
$\mu/\mu_0$ at the LHC. }
\end{center}
\end{figure}

In Fig.\ref{fig3} we present the integrated LO and the QCD NLO
corrected cross sections for the \ppWWbb processes at the LHC, as the
functions of the renormalization/factorization scale. The full curve
is for the LO cross section. The dashed and dotted curves are for
the QCD NLO corrected cross sections by taking the QCD NLO
contribution collection scheme-I, and adopting the inclusive and
exclusive two-jet event selection schemes, separately. The
dash-dot-dotted and dash-dotted curves are for the results by
taking the QCD NLO contribution collection scheme-II, and adopting
the inclusive and exclusive two-jet event selection schemes,
respectively.

\par
We can see from Fig.\ref{fig3} that the curve for the LO cross
section is relatively smooth, while the QCD NLO corrections in the
QCD NLO contribution collection scheme-I make the uncertainty on
$\mu$ rather large. The QCD NLO corrections in the QCD NLO
contribution collection scheme-I by using the inclusive
("exclusive") two-jet event selection scheme change the LO cross
section by a factor of between $1.30$ ($0.51$) and $1.99$ ($1.60$) when
$\mu$ varies from $0.1\mu_0$ to $3\mu_0$. In other words, the scale
uncertainties of the QCD NLO corrected cross sections in the QCD NLO
contribution collection scheme-I are even worse than that of the LO
cross section. This enhancement of the uncertainty of the QCD NLO
theoretical predictions is partially due to using the 5FS in
calculations, and in this scheme the perturbative series is not
ordered strictly by powers of the $\alpha_s$. Figure \ref{fig3} shows
also that the QCD NLO corrections in the QCD NLO contribution
collection scheme-II by adopting the inclusive ("exclusive")
two-jet event selection scheme increase the LO cross section by a
factor varying from $2.12$ ($1.59$) to $3.54$ ($2.68$) in the
plotted range of $\mu$. That shows the QCD corrections in the QCD
NLO contribution collection scheme-II violate the convergence of the
perturbative series in some ranges of $\mu$, particularly the range
of $\mu < 0.5\mu_0$. But we find that the total cross section for
the $W^+W^-b(\bar{b})$ production up to the QCD NLO in the QCD NLO
contribution collection scheme-II demonstrates a weaker scale
uncertainty compared with the LO one. That means the energy scale
uncertainty of the LO cross section can be improved by taking the
QCD NLO contribution collection scheme-II and adopting either the
inclusive or the exclusive two-jet event selection schemes. It
just reflects that a stable prediction of the integrated cross
section for the \ppWWbb process requires not only a veto on a second
isolated hard jet, but also the inclusion of the QCD NLO corrections
contributed by the $W^+W^-b\bar{b}(bb, \bar{b}\bar{b})$ production
events with the two final $b(\bar{b})$ quarks being merged as one
jet. In the further numerical calculations, we take $\mu=\mu_0$ in
default of other statements.

\vskip 5mm
\subsection{Dependence on Higgs mass  }
\par
As indicated in the introduction, the $W^+W^-b(\bar{b})$ production
may serve as one of the important backgrounds of Higgs-boson
production associated with a $b(\bar{b})$ jet, if the Higgs-boson
mass is larger than $2m_W$. In this subsection we study the behavior
of the QCD NLO corrected cross section for the \ppWWb process as a
function of the Higgs mass. According to the recent Fermilab
report\cite{higgs_mass}, the SM Higgs-boson mass has been narrowed
down to $114 \sim 158~GeV~$ and $175 \sim 185~GeV$. In our numerical
calculations we choose Higgs mass values and their corresponding
widths being $\Gamma_H(m_H=120~GeV)=0.2965\times 10^{-2}~GeV$,
$\Gamma_H(m_H=150~GeV)=1.704\times 10^{-2}~GeV$ and
$\Gamma_H(m_H=180~GeV)=0.6511~GeV$ separately, as provided in
Sec.\ref{parameters}. In Table \ref{tab1} we present the numerical
results of the LO and QCD NLO corrected cross sections, and the
corresponding K factors by taking the QCD NLO contribution
collection scheme-I with the Higgs-boson mass being $120$, $150$ and
$180~GeV$, respectively. The notations of (inc.) and (exc.) refer to
the results by adopting the inclusive and exclusive two-jet
event selection schemes, separately. From this table we find that
the LO and QCD NLO corrected cross sections with both the two
two-jet event selection schemes are insensitive to the Higgs-boson
mass when $m_H$ varies from $120$ to $180~GeV$. That is because the
contributions of the Feynman diagrams, which correspond to the $H^0
b$ production with $H^0$ subsequently decaying into $W$ pair, are
very small.
\begin{table}
\begin{center}
\begin{tabular}{|c|c|c|c|c|c|}
\hline $m_{H}$(GeV) & $\sigma_{LO}$ (pb) & $\sigma_{NLO}$ (inc.)(pb)
& K factor & $\sigma_{NLO}$ (exc.)(pb) & K factor \\
\hline 120 & 61.86(2) & 102.84(4) & 1.66 & 74.56(2) & 1.21 \\
\hline 150 & 61.84(2) & 102.82(4) & 1.66 & 74.56(2) & 1.21 \\
\hline 180 & 61.96(2) & 102.58(4) & 1.67 & 75.18(2) & 1.21 \\
\hline
\end{tabular}
\end{center}
\begin{center}
\begin{minipage}{15cm}
\caption{\label{tab1} The integrated LO and QCD NLO corrected cross
sections and the corresponding K factors for the $pp \to W^+W^-b+X$
and $pp \to W^+W^-\bar{b}+X$ processes at the LHC by taking the QCD
NLO contribution collection scheme-I. }
\end{minipage}
\end{center}
\end{table}

\vskip 5mm
\subsection{$p_{T,b}$ and $p_{T,W^\pm}$ distributions }
\par
\begin{figure}
\begin{center}
\includegraphics[scale=1.0]{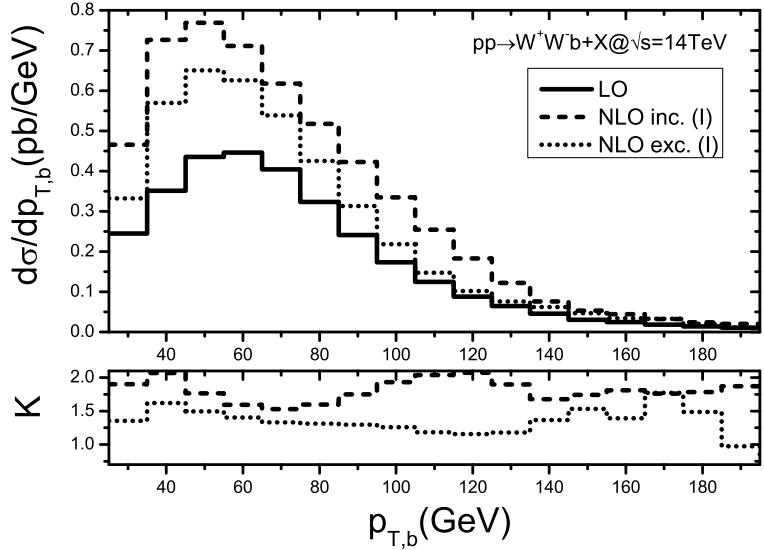}
\hspace{0in}%
\caption{\label{fig4} The LO, QCD NLO corrected distributions of the
transverse momentum of the bottom jet ($p_{T,b}$) and corresponding
K factors for the \ppWWb process at the LHC by taking $\mu = \mu_0$,
$m_H = 120~GeV$ and adopting the QCD NLO contribution collection
scheme-I. The corresponding K factors are shown in the figure below.
}
\end{center}
\end{figure}

The LO and QCD NLO corrected distributions of the transverse
momentum of the $b$ jet ($p_{T,b}$) and the corresponding K-factors for
the \ppWWb process at the LHC by taking the QCD NLO contribution
collection scheme-I, are demonstrated in Fig.\ref{fig4}. The full,
dashed and dotted curves are for the LO, QCD NLO corrected
$p_{T,b}$ distributions by adopting the inclusive and exclusive
two-jet event selection schemes, separately. The corresponding
K factors are drawn in the attached figure below. These two figures
show that the NLO QCD corrections with both the inclusive and
exclusive two-jet event selection schemes always enhance the LO
differential cross section $d \sigma_{LO}/d p_{T,b}$, and the NLO
QCD corrections to the LO differential cross section with the
inclusive scheme are always larger than that with the exclusive
scheme.

\par
\begin{figure}[htbp]
\includegraphics[scale=0.75]{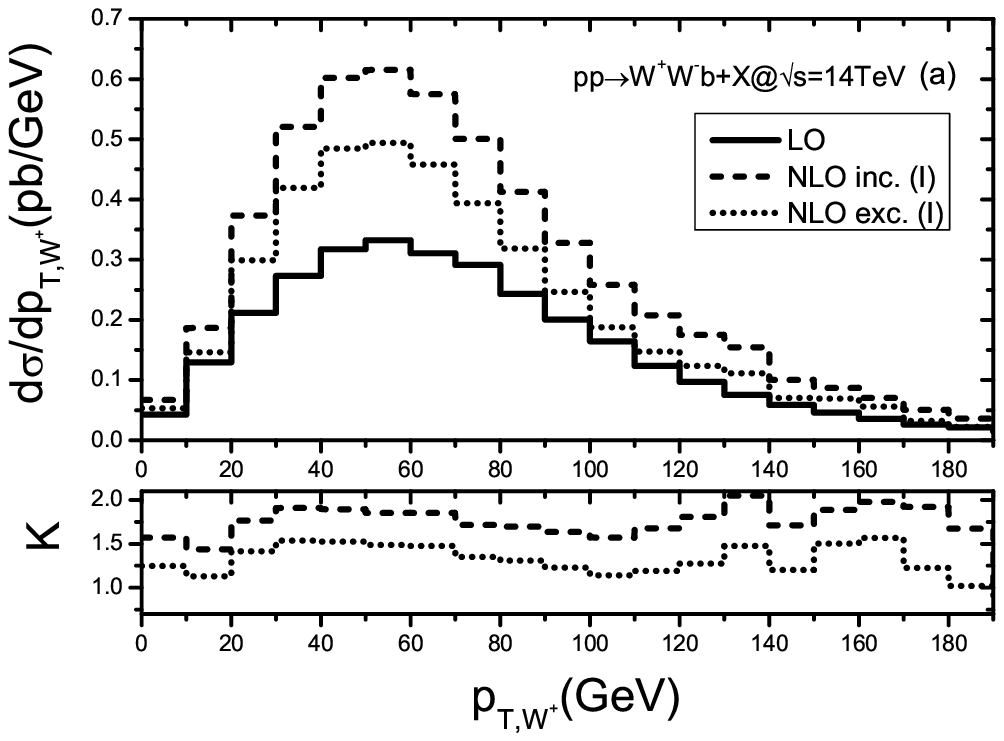}
\includegraphics[scale=0.75]{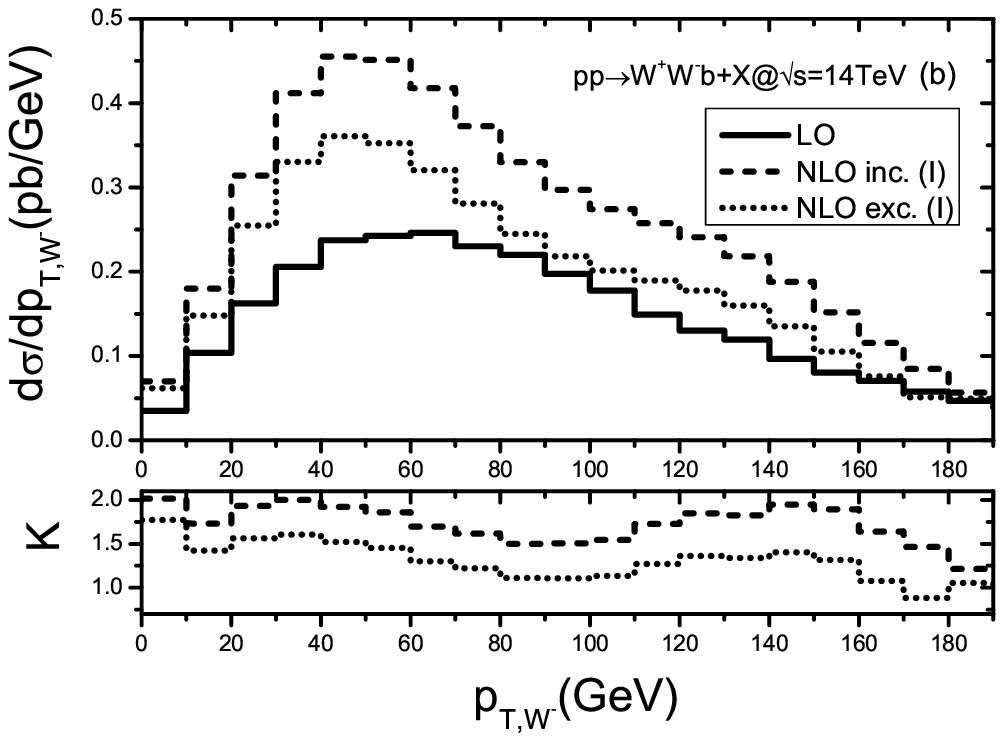}
\hspace{0in}%
\caption{\label{fig5} (a) The LO, QCD NLO corrected distributions of
the transverse momentum of $W^+$ boson and the corresponding
K factors for the \ppWWb process at the LHC. (b) The LO, NLO QCD
corrected distributions of the transverse momentum of $W^-$ boson
and the corresponding K factors. In these two figures we take
$\mu=\mu_0$, $m_H=120~GeV$ and adopt the QCD NLO contribution collection scheme-I. }
\end{figure}

We plot the LO and the QCD NLO corrected distributions of the
transverse momenta of $W^{\pm}$ bosons $p_{T,W^+}$ and $p_{T,W^-}$
for the \ppWWb process at the LHC by taking the QCD NLO contribution
collection scheme-I in Fig.\ref{fig5}(a) and Fig.\ref{fig5}(b),
respectively. We use the full, dashed and dotted curves describing
the LO, QCD NLO corrected distributions with the inclusive and
exclusive two-jet event selection schemes, separately. We can see
from the figures that the QCD NLO corrections with both the
inclusive and exclusive two-jet event selection schemes, always
increase the LO differential cross sections $d \sigma_{LO}/d
p_{T,W^+}$ and $d \sigma_{LO}/d p_{T,W^-}$, and the enhancement due
to the QCD NLO corrections to $d \sigma_{LO}/d p_{T,W}$ with the
inclusive two-jet event selection scheme is larger than that with
the exclusive two-jet event selection scheme. From the figures we
see also that the distributions of the transverse momenta of $W^+$
and $W^-$ bosons are different.

\vskip 5mm
\subsection{Invariant mass $m_{(W^+b)}$ distribution }
\par
\begin{figure}
\begin{center}
\includegraphics[scale=1.0]{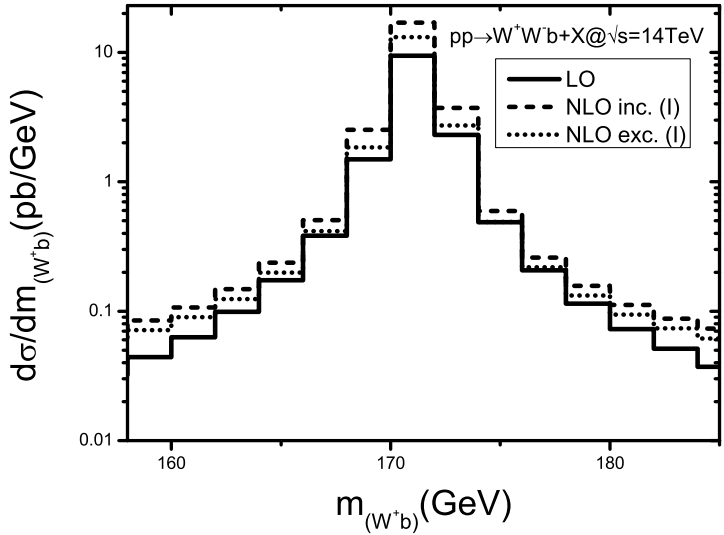}
\hspace{0in}%
\caption{\label{fig6} The distributions of the invariant mass of the
bottom jet and $W^+$ boson, $m_{(W^+b)}$, for the \ppWWb process at
the LHC, where we take $\mu=\mu_0$, $m_H=120~GeV$ and adopt the QCD
NLO contribution collection scheme-I. }
\end{center}
\end{figure}

We plot the distributions of the invariant mass of the $(W^+b)$ pair,
$m_{(W^+b)}$, at the LO and the QCD NLO in Fig.\ref{fig6} with
$\mu=\mu_0$ and $m_H=120~GeV$, where the NLO distributions of
$m_{(W^+b)}$ are presented by taking the QCD NLO contribution
collection scheme-I and adopting the inclusive and exclusive
two-jet event selection schemes separately. We can see from the
figure that the LO and QCD NLO corrected differential cross
sections, $d \sigma_{LO}/dm_{(W^+b)}$ and $d
\sigma_{NLO}/dm_{(W^+b)}$, have peaks in the vicinity of
$m_{(W^+b)}= m_t$. It demonstrates that there is a large
contribution part for the process \ppWWb, which comes from the
single top production associated with a $W^-$ boson ($pp \to bg \to
W^- t$) followed by the subsequential decay of the top quark to $W^+
b$. That is to say the main contribution to the \ppWWb process
originates from the Feynman diagrams of Figs.\ref{fig1}(4,6) at the
LO, and their related one-loop diagrams at the NLO. Here we can see
that in the invariant mass range around $m_{(W^+b)}\sim 171~GeV$ the
QCD NLO corrections with both the inclusive and exclusive
two-jet event selection schemes strongly enhance the LO invariant
mass distribution $d\sigma_{LO}/dm_{(W^+b)}$, and the QCD NLO
correction with the inclusive scheme enhances the LO distribution
more heavily than that with the exclusive scheme, particularly
around the resonance peak.

\vskip 5mm
\subsection{Cross sections for the $pp \to W^+ W^- b \bar{b},~W^+ W^- b b,
~W^+ W^- \bar{b} \bar{b}$  processes }
\par
In Table \ref{tab2} we list the numerical results of the integrated
cross sections for the three processes $pp \to W^+W^-b\bar{b}$ and
$pp \to W^+W^-bb,~\bar{b}\bar{b}$ with different Higgs-boson mass
values. These three $W^+W^-b\bar{b}$, $W^+W^-bb$ and
$W^+W^-\bar{b}\bar{b}$ production processes are induced by the
$gg(q\bar q) \to W^+W^-b\bar b$, $bb \to W^+W^-bb$ and $\bar b \bar
b \to W^+W^-\bar b\bar b$ partonic processes separately. The data in
columns (I), (II) and (III) are the integrated cross sections over
three different phase-space regions, respectively.

\par
In phase-space region (I), the two $b(\bar{b})$ jets are
isolated (i.e., the two $b(\bar{b})$ jets satisfy $\sqrt{\Delta
\eta^2 + \Delta \phi^2} > 0.7$), and one $b(\bar{b})$ jet is
detectable satisfying the constraints of $p_{T,b(\bar{b})} > 25~GeV$
and $|y_{b(\bar{b})}| < 2.5$ while the other $b(\bar{b})$ jet is
undetectable, i.e., satisfying $p_{T,b(\bar{b})} < 25~GeV$ or
$|y_{b(\bar{b})}| > 2.5$.

\par
In phase-space region (II), the two $b(\bar{b})$ jets are merged
as one jet (i.e., $\sqrt{\Delta \eta^2 + \Delta \phi^2} < 0.7$), and
the merged jet passes the constraints of $p_{T,j} > 25~GeV$ and
$|y_{j}| < 2.5$. That is just the additional QCD NLO contribution
when the QCD NLO contribution collection scheme-II is adopted.

\par
In phase-space region (III), the two $b(\bar{b})$ jets are
separately detectable. This means that $\sqrt{\Delta \eta^2 + \Delta
\phi^2} > 0.7$ and the two $b(\bar{b})$ jets separately satisfy the
constraints of $p_{T,b(\bar{b})} > 25~GeV$ and $|y_{b(\bar{b})}| <
2.5$.

\par
Analogy of the data in Table \ref{tab1}, Table \ref{tab2} shows that the
integrated cross sections for the processes $pp \to
W^+W^-b\bar{b}(bb,\bar{b}\bar{b})$ over phase-space regions (I),
(II) and (III) are insensitive to the Higgs-boson mass. That means
the contributions from the Feynman diagrams, which correspond to the
$H^0 b$ production followed by the decay of $H^0 \to W^+W^-$, are
rather small. We also find from Table \ref{tab2} that the
integrated cross sections for the $pp \to W^+W^-b\bar{b}$, $pp \to
W^+W^-bb$ and $pp \to W^+W^-\bar{b}\bar{b}$ processes over the three
phase-space regions are all rather large, particularly phase-space 
region (III). Our calculation demonstrates that the contribution from the
$pp \to W^+W^-b\bar b+X$ process is the largest one, which has a $2\sim 3$
order larger contribution to the cross section than the other two processes.
The large enhancement of this process with the final $b\bar{b}$ pair is due
to the large gluon luminosity in PDF at the LHC and the overcompensation
of the top quark resonances in the $pp \to t\bar t \to W^+W^-b\bar b+X$ process.
\begin{table}
\begin{center}
\begin{tabular}{|c|c|c|c|}
\hline $m_{H}$(GeV) &  $WWb\bar{b}(bb,\bar{b}\bar{b})$(I)(pb) &
$WWb\bar{b}(bb,\bar{b}\bar{b})$(II)(pb)
& $WWb\bar{b}(bb,\bar{b}\bar{b})$(III)(pb)\\
\hline 120 &  141.66(6) & 27.15(2) & 603.2(1) \\
\hline 150 &  141.54(6) & 27.15(2) & 603.5(1) \\
\hline 180 &  141.28(6) & 27.15(2) & 603.3(1) \\
\hline
\end{tabular}
\end{center}
\begin{center}
\begin{minipage}{15cm}
\caption{\label{tab2} The integrated cross sections for the three
processes $pp \to W^+ W^- b \bar{b},~W^+ W^- b b,~W^+ W^- \bar{b}
\bar{b}$ over three different phase-space regions (I), (II) and
(III) with $\mu=\mu_0$, $m_H = 120$, $150$ and $180~GeV$,
respectively. }
\end{minipage}
\end{center}
\end{table}

\vskip 5mm
\section{Summary}
\par
In this paper we calculate the full QCD NLO corrections to the
$W$-pair production in association with a massive (anti)bottom jet
in the SM at the LHC, which is an important background of the single
top production and can be used to search for new physics beyond the
SM by using the five-flavor scheme. The dependence of the integrated
cross sections on the factorization/renormalization scale and the Higgs
mass is studied. We also investigate the QCD NLO corrections to the
distributions of the transverse momenta ($\frac{d\sigma}{dp_T}$) of
the final particles. Our numerical results show that by taking the
QCD NLO contribution collection scheme-I and adopting the
inclusive (exclusive) two-jet event selection scheme, the K factor of the
total cross section for the \ppWWbb process at the LHC can reach
$1.66$ ($1.21$) with $\mu=\mu_0$. We find that the LO integrated and
differential cross sections are modified by the QCD NLO radiative
corrections obviously, and the QCD NLO corrections to the
$W^+W^-b(\bar{b})$ production process with the QCD NLO contribution
collection scheme-I can not reduce the scale uncertainty of the LO
cross section for the \ppWWbb process at the LHC by adopting either
the inclusive or exclusive two-jet event selection schemes. The
stable prediction for the integrated cross section requires not only
a veto on a second isolated hard jet, but also the inclusion of the
QCD NLO correction contributed by the $W^+W^-b\bar{b}(bb,
\bar{b}\bar{b})$ production events with the two final
$b(\bar{b})$ quarks being merged as one jet.

\vskip 5mm
\par
\noindent{\large\bf Acknowledgments:} This work was supported in
part by the National Natural Science Foundation of
China (Contract No.10875112, No.11075150, No.11005101), and the Specialized
Research Fund for the Doctoral Program of Higher
Education (Contract No.20093402110030).


\vskip 10mm
\begin{thebibliography}{99}
\bibitem{Tait}
  T. M. P. Tait, Phys. Rev. {\bf D 61}, 034001(1999); J. Alwall et al.,
      Eur. Phys. J. C. {\bf 49}, 791 (2007); J. A. Aguilar-Saavedra and
      A. Onofre, Phys. Rev. {\bf D 83}, 073003 (2011).
\bibitem{3}
  Tevatron Electroweak Working Group for CDF and D0, [arXiv:0908.2171].

\bibitem{4}
  D0 Collaboration, Phys. Rev. Lett. 103 (2009) 092001 , [arXiv:0903.0850];
  D0 Collaboration, Phys.Lett.{\bf B682} (2010) 363 ,[arXiv:0907.4259];
   CDF Collaboration,Phys.Rev.Lett.103 (2009) 092002  , [arXiv:0903.0885].

\bibitem{Carlson}
  Carlson D. O. and Yuan C. P., Phys. Lett. {\bf B306} (1993) 386;
  Mahlon G. and Parke S. J., Phys. Rev. {\bf D55} (1997) 7249, [arXiv:hep-ph/9611367];
  Heinson A. P., Belyaev A. S. and Boos E. E., Phys. Rev. {\bf D56} (1997) 3114, [arXiv:hep-ph/9612424];
  Mahlon G. and Parke S. J., Phys. Lett. {\bf B476} (2000) 323, [arXiv:hep-ph/9912458];
  Sullivan Z., Phys. Rev. {\bf D72} (2005) 094034,[arXiv:hep-ph/0510224].

\bibitem{Gerber}
 C. E. Gerber et al., Report No. FERMILAB-CONF-07-052, (2007),
       and references therein (unpublished).
\bibitem{Mellado}
  B. Mellado, W, Quayle, Sau Lan Wu, TeV4LHC workshop, 29 April
  2005; B. Mellado, W. Quayle and S. L. Wu, Phys. Rev. {\bf D76}, 093007 (2007),
  [arXiv:0708.2507].

\bibitem{wwjet_0908.4124}
  Stefan Dittmaier, Stefan Kallweit, and Peter
  Uwer, Nucl. Phys. {\bf B826} (2010) 18-70, [arxiv:0908.4124]; 'NLO QCD
  corrections to WW+jet production including leptonic W decays at
  hadron colliders', FR-PHENO-2010-002, PSI-PR-10-04, HU-EP-10/02,
  [arxiv 1001.2427].

\bibitem{wwjet 0710.1832}
  John M. Campbell,R. Keith Ellis, Giulia Zanderighi, JHEP 0712 (2007)
  056, [arxiv:0710.1832].

\bibitem{fey}
  T. Hahn,  Comput. Phys. Commun. {\bf 140} (2001) 418.

\bibitem{formloop}
  T. Hahn, M. Perez-Victoria, Comput. Phys. Commun. {\bf 118}
  (1999) 153.

\bibitem{CMS}
  A. Denner, S. Dittmaier, M. Roth, D. Wackeroth, Nucl. Phys. {\bf
  B560} (1999) 33; A. Denner, S. Dittmaier, M. Roth, L.H. Wieders,
  Nucl. Phys. {\bf B724} (2005) 247.

\bibitem{je_algo}
  S.D. Ellis and D.E. Soper, Phys. Rev. {\bf D48} (1993) 3160, [arXiv:hep-ph/9305266].

\bibitem{19}
  B. W. Harris and J. F. Owens,  Phys. Rev. {\bf D65} (2002) 094032,
  [arxiv:hep-ph/0102128].

\bibitem{gs}
  J. Collins, F. Wilczek, and A. Zee, Phys. Rev. {\bf D 18}, 242 (1978);
       W. J. Marciano, Phys. Rev. {\bf D 29}, 580 (1984); P. Nason, S. Dawson,
       R.K. Ellis, Nucl. Phys. {\bf B 327}, 49 (1989); Nucl. Phys. {\bf B 335},
       260(E) (1989).

\bibitem{Stefan}
 R. K. Ellis, G. Zanderighi, JHEP 0802 (2008) 002, [arXiv:0712.1851].

\bibitem{OneTwoThree}
  G.'t Hooft and M. Veltman, Nucl. Phys. {\bf B153} (1979) 365.

\bibitem{Four}
  A. Denner, U. Nierste and R. Scharf, Nucl. Phys. {\bf B367} (1991) 637.

\bibitem{Five}
  A. Denner and S. Dittmaier, Nucl. Phys. {\bf B658} (2003) 175.

\bibitem{ff}
  G. J. van Oldenborgh, Comput. Phys. Commun. {\bf 66}, 1 (1991).

\bibitem{guolei}
  G. P. Lepage, J. Comput. Phys. 27 (1978)  192 and CLNS-80/447.

\bibitem{ini_coll}
J. Collins Phys.Rev. {\bf D58} (1998) 094002.


\bibitem{FFS}
  B. A. Kniehl, G. Kr\"amer, I. Schienbein, and H. Spiesberger,
        Eur. Phys. J. {\bf C 41}, 199 (2005); S. Kretzer and I. Schienbein,
        Phys. Rev. {\bf D 58}, 094035 (1998); M. Kr\"amer, F. Olness, and
        D. Soper, Phys. Rev. {\bf D 62}, 096007 (2000); F. Olness, R. Scalise and
        W-K. Tung, Phys. Rev. {\bf D 59}, 014506 (1998).

\bibitem{hepdata}
  C. Amsler,{\it et al.} Phys. Lett. {\bf B667} (2008) 1 .

\bibitem{cteq}
  J. Pumplin \textit{et al}., JHEP 0207 (2002) 012 ;
  D. Stump \textit{et al}., JHEP 0310 (2003) 046 .

\bibitem{twidth}
  M. Je\.zabek and J.H. K\"uhn, Nucl. Phys. {\bf B314} (1989) 1.

\bibitem{hdecay}
  A. Djouadi, J. Kalinowski, M. Spira, Comput. Phys. Commun. 108
  (1998) 56.

\bibitem{higgs_mass}
  The CDF Collaboration, the D0 Collaboration, the Tevatron New Physics,
        and Higgs Working Group, Fermilab, Report No. FERMILAB-CONF-10-257-E, (2010), [arXiv:1007.4587v1].

\end{thebibliography}
\end{document}